# The Strange Metal State of the high-Tc Cuprates

Richard L. Greene

Quantum Materials Center and Department of Physics

University of Maryland, College Park, MD 20742

Abstract

This paper is dedicated to the memory of Professor K. Alex Müller. I present a few remarks about my interactions with Alex over the years. Then I present a very brief summary of recent transport studies of the strange metal normal-state in the high-temperature copper oxide superconductors, a subject of great interest to Alex.

Introduction

I first met Alex sometime in early 1982 when he came for a visit to the IBM San Jose (now Almaden) Research Laboratory. I was the leader of a group that had just discovered the first example of superconductivity in a two-dimensional organic metal [1] and Alex was quite interested in this new result, as well as other research that I was doing on unconventional superconducting materials. Although I was unfamiliar with the details of Alex's research on phase transitions, critical phenomena and ferroelectrics, he was well known to me as one of the few IBM Fellows (the most prestigious individual honor at IBM). I also knew that he had recently spent two years at the IBM Yorktown Heights laboratory where, among other projects, he did some very nice work on granular superconductors. His time at Yorktown (1978-80) marked the beginning of his interest and research in superconductivity. After this, I had occasional email interactions with Alex about my research on organic superconductors. When he learned in early 1985 that I was planning to take a one year sabbatical, he invited me to come to the IBM Rüschlikon laboratory to work with him on some oxide materials that he thought might be interesting superconductors. This was a very attractive idea for me, both because of the science and because I wanted to go skiing with Alex (another activity for which he was quite famous). To my great regret in hindsight, I decided to spend my sabbatical year (85-86) at IBM Yorktown instead!

In late August 1986, just after I had moved permanently to IBM Yorktown, I saw an internal research report from Alex and Georg Bednorz at IBM Rüschlikon on the possible superconductivity in the La-Ba-Cu-O system (soon to be published in Z. Phys. [2]). I was very excited about this report and I immediately called Alex and asked him to send me some samples so that I could confirm his results with critical field and specific heat measurements. He agreed, and thus began my collaboration with Alex and Georg that resulted in the only contributed talk on the high-Tc copper oxides (cuprates) at the 1987 APS March meeting in New York (see Fig. 1). Most of this work was never published because the La-Ba-Cu-O polycrystalline samples were not single phase and no specific jump at Tc could be observed. However, by mid-October I had found that the resistivity drop in these samples was shifted to lower temperature by application of a magnet field and Alex and Georg had found a small diamagnetic susceptibility below 30K. These two results convinced me that superconductivity in the La-Ba-Cu-O system was indeed real and represented a seminal scientific breakthrough. Surprisingly, in spite of Alex's outstanding scientific reputation, his internal report and magnetic susceptibility results [3] generated very little interest at IBM Yorktown.  This all dramatically changed in early December 1986 when the Tokyo group presented their results at a MRS meeting in Boston, results which confirmed the Rüschlikon work in single phase samples [4]. After that the worldwide frenzy of research on cuprates began, which quickly led to Paul Chu and M.K. Wu's discovery of YBCO with a Tc above liquid nitrogen temperatures!! Within three months there were so many new experimental results on cuprates that the famous "Woodstock of Physics" night session at the 1987 APS March meeting was arranged--an unprecedented event at a March meeting. When I presented my contributed talk the day before the famous night session I introduced Alex to the overflowing audience and he received a thunderous ovation. Very few of the superconductor "experts" in the audience knew Alex, but the significance of his cuprate superconductivity discovery was recognized by all!!

The discovery of high-Tc superconductivity in the copper oxides (cuprates) was significant in many ways. It not only changed the accepted ideas about the limits of Tc in superconductors at ambient pressure but it changed the thinking about the role of strong electron correlations in the electronic properties of materials. Many new and interesting strongly correlated superconductors have been discovered since 1987 but none have an ambient pressure Tc as high as the cuprates, nor was their discovery as momentous as that of the cuprates. In spite of the over 100,000 experiments carried out on cuprates since 1987 (and much theoretical work), the microscopic mechanism that causes the superconductivity is still not universally agreed upon, although a recent experiment [5] strongly confirms the view that charge-transfer superexchange is the key to the electron pairing.  Moreover, the normal state from which the superconductivity emerges is very unconventional (i.e., non-Fermi liquid) and most of the normal state electronic properties are quite anomalous and still unexplained. Alex was very interested in experiments related to the non-Fermi liquid nature of the normal-state of the cuprates. In fact, this was the topic of my last in-person meeting with Alex at the 2015 M2S conference in Geneva (see Fig. 2).

Transport in the strange metal

In this paper I will present a few brief comments on the experimental status of the so-called "strange metal" normal-state of the cuprates. Most of what I will say here is discussed in more depth by several review articles [6-8]. The first indication that the normal-state was anomalous was the 1989 observation of a T-linear resistivity from 7-700K in a low Tc Bi2201 cuprate [9].

The resistivity had the same linear slope at all temperatures, i.e., did not saturate at the highest temperature and violated the Mott-Ioffe-Regal (MIR) limit. This anomalous high-temperature resistivity behavior has been called a bad metal while the low temperature linear behavior has been called a strange metal. Here I will only discuss the strange metal behavior. The lower temperature T-linear resistivity has now been seen in many other superconducting cuprates that are driven into the normal-state by a magnetic field [10]. That this behavior is not caused by electron-phonon scattering in a low carrier density material was proven by the T-linear resistivity seen in the electron-doped cuprates down to 35 mK [11]. So, the ground state of the superconducting cuprates is certainly not that of a Fermi liquid! On the other hand, the non-superconducting cuprates found at higher doping are Fermi liquids. The early very low temperature T-linear resistivity results were found at only one doping and were interpreted as arising from a quantum critical point [12]. However, later data has shown that the T-linear behavior is found over a wide range of doping in both hole- and electron-doped cuprates [13,14], so it appears quantum criticality is not the explanation [see 15 for a different theoretical model].

Another conjecture for the origin of low temperature T-linear resistivity is based upon the Planckian dissipation time $t_{Pl} = \hbar/kT$, the shortest time for energy loss in a many-body quantum system) [16]. If one extracts a scattering rate from the resistivity [17] or ADMR [18] using a standard Fermi liquid approach one finds in many cuprates a very close agreement with the Planckian scattering rate, i.e. $\hbar/\tau_{Pl} = \alpha\, kT$, with $\alpha \sim 1$. However, the Planckian origin of T-linear resistivity has been challenged by many [19-21] and Planckian dissipation remains only an interesting conjecture at this time.

More significantly, the slope (strength) of the T-linear resistivity has been shown to be directly correlated with the magnitude of Tc [14, 22-24]! An example of this correlation for an electron-doped (n-type) cuprate is shown in Fig. 3[23]. Other work has shown that the superfluid density is correlated with the T-linear resistivity slope in a hole-doped (p-type) cuprate [25]. And, anomalous behavior in magnetoresistance and thermopower (discussed below) is also correlated with Tc. All of these results strongly suggest that whatever causes the T-linear resistivity is also responsible for the electron pairing in the superconducting state.

There are a few other puzzling aspects to the resistivity in the cuprates that are worth mentioning here. In the hole-doped cuprates, the low temperature T-linear behavior continues to very high temperatures (apparently through the MIR limit) with NO change in slope. This suggests that the cause of the T-linear resistivity is the same at all temperatures in hole-doped cuprates. In contrast, in electron-doped cuprates the low temperature T-linear resistivity becomes approximately quadratic above ~50K and continues as ~ $T^2$ up to the highest temperatures that can be measured [26]. This is clearly a different kind of strange metal behavior since $T^2$ resistivity is not conventional at temperature above ~ $\theta_D/5$. Why the p-type and n-type cuprates have such different resistivity behavior above ~ 50K is presently unexplained. Moreover, if one extracts a scattering rate from the resistivity above 50K in n-type cuprates one finds that it exceeds the Planckian limit [21]!!

Another strange metal transport behavior that has been studied in detail more recently is the magnetoresistance (MR). In both n- and p-type cuprates an unconventional (non-Fermi liquid) H-linear MR has been found that appears to be correlated with the T-linear resistivity [27-29]. Like the T-linear resistivity, the H-linear MR is found over a range of doping, in particular for doping above p* (the end point of the pseudogap in hole-doped cuprates and the end of the antiferromagnetic phase in the electron-doped cuprates). So, this is not a manifestation of something quantum critical unless there is an extended critical region for some unknown reason. An example of the linear MR and its temperature dependence is shown in Fig.4. There are several aspects of this data which are "strange", i.e. not compatible with standard Boltzmann transport theory. First, the H-linear behavior of the MR is found when $\omega_c \tau \ll 1$ where an $H^2$ dependence is expected. Second, the magnitude of the H-linear MR has no temperature dependence below some temperature. In Boltzmann theory the T dependence of the MR is often found to obey Kohler scaling. In the case of the cuprates Kohler scaling does not work (except in the overdoped Fermi liquid state) but another empirical H/T scaling does fit the data. This H/T scaling is slightly different for n-type and p-type cuprates [28,29] but for both the transport is dominated by the strongest energy scale, either kT or uH. The reason for the subtle difference between n- and p-type scaling (one quadratic, one linear) is not known. For completeness one should note that at one doping (p =0.24) of hole-doped Nd-LSCO the standard Boltzmann theory does appear to work [30]. But, for all other hole-doped cuprates so far measured the standard theory does not work [29]!! This difference is not yet understood but may result from the unique situation in Nd-LSCO where the doping p =0.24 is close to a Lifshitz transition of the Fermi surface.

The low temperature thermopower is another transport property that is strange. An example is shown in Fig. 5 for a superconducting (at H =0) electron-doped cuprate, $La_{2-x}Ce_xCuO_4$ (LCCO), with p > p* and H > Hc2. The log(1/T) increase of S/T as temperature decreases is in marked contrast to the Fermi liquid expectation of a constant S/T with decreasing temperature. The logarithmic S/T is that expected at a quantum critical point, however, in LCCO this behavior is found at all doping from p* to the end of the SC dome [31]. The same logarithmic S/T behavior is found in another n-type cuprate system, PCCO [31]. In addition, for both n-type systems the magnitude of the logarithmic S/T decreases with increasing doping in apparent correlation with the decrease of Tc with doping. Thus, in electron-doped cuprates the magnitudes of the strange metal MR, resistivity, and thermopower decrease as Tc decreases for p > p*[28]. This suggests a correlation between the low temperature strange metal normal state and the cause of the unconventional SC in the cuprates. For hole-doped cuprates a low temperature normal state S/T ~ log (1/T) has been found for Nd-LSCO at p*[32]. However, except for this one result, there are not yet sufficient low temperature normal state thermopower measurements to know if a logarithmic S/T is found over a range of doping in Nd-LSCO and in other hole-doped cuprates. There is no reason to expect anything different in hole-doped systems. In fact, as far as transport data is concerned, the overdoped (p >p*) strange metal normal state and the superconducting state of the electron- and hole-doped cuprates are basically the same.

Another transport property that has been measured extensively in the cuprates is the Hall effect. The temperature and doping dependence of the Hall coefficient $R_H$ was found to have unconventional behavior. For example, in the single band hole-doped cuprate LSCO at optimal doping $R_H$ varies as 1/T in violation of Fermi-liquid theory [33]. Moreover, at low temperature $n_H(0)$ increases anomalously from p to 1+p over a wide range of doping[34]. Similar strange metal Hall behavior is found in many other hole-doped cuprates [35] and in electron-doped cuprates [12] (although the presense of two bands near optimal doping complicates the interpretation here). In addition, the inverse Hall angle has a strange (non-FL) temperature dependence in all cuprates. Unfortunately, it is not possible to cite all the interesting experimental work done on the Hall effect in the cuprates in this short review. Most of the experimental Hall effect papers have been interpreted using conventional Boltzmann transport theory even though it is clear that the resistivity and MR cannot be explained by conventional theory. This can be forgiven since there is no other theoretical framework to explain Hall data at this time. One exception to this is a recent paper from Shekhter et al. [36] where the authors attempt to explain the Hall data in LSCO via a strange metal phenomenology similar to that used to explain their earlier MR data [27]. There have also been many theoretical papers devoted to an explanation of the Hall effect in cuprates. Many of these use some version of a strongly correlated Hubbard model [37]. But, other approaches also exist [38]. Needless to say, this reference list is not exhaustive and at the present time there is no agreed upon explanation of the Hall effect in the cuprates.

A few more comments. The frequency and temperature dependence of the optical properties of the cuprates are also rather unconventional [39,40]. I do not have the space (or expertise) to discuss them here but they will be important input for a complete understanding of the strange metal state in the cuprates. There is no presently accepted theory for the strange metal state, but there is universal agreement of its importance in reaching a complete understanding of strongly correlated metals and the unconventional superconductivity found in such materials. I list here a few of the recent theoretical papers that I have seen [41-44] which include reference to earlier theoretical work. This is far from a complete list and I apologize to those many colleagues and friends omitted.

Concluding remarks

In this short paper dedicated to the memory of K. Alex Müller I have tried to highlight the most recent experiments on the transport properties of the strange metal normal-state of the cuprates. And the many puzzles that they present. Alex was as puzzled as his peers about the unconventional properties of the cuprates, although he certainly had his strong opinions about what was causing them. But, I'm certain that even the great scientist that Alex was in 1986 could not imagine how his pursuit of Jahn-Teller polarons as a driver of higher Tc superconductivity would lead to the astounding new physics that has been discovered in oxides since then. All of us who worked with Alex will miss his humble nature and brilliant scientific mind. Personally, I am sad that I never made the time to swoosh down a ski slope with Alex to see his unbridled joy at being at home in the mountains.

Acknowledgements

I thank the following colleagues for fruitful discussions on this topic over many years: Nigel Hussey, Louis Taillefer, Sankar Das Sarma, Nick Poniatowski, Aharon Kapitulnik, Tara Sarkar, Nick Butch, Johnpierre Paglione, Yoran Dagan, Patrick Fournier, Kui Jin, Peter Armitage, Chandra Varma, Greg Boebinger, Andrey Chubukov, Steve Kivelson, and others that my old mind has forgotten. My experimental research has been supported for many years by the National Science Foundation, at present under award #2002658.

SESSION FN: SUPERCONDUCTIVITY—GENERAL
Tuesday morning, 17 March 1987
Nassau A Room at 11:00
M. Tinkham, presiding

11:36
FN 4 Specific Heat of BaLaCuO Superconductors. R. L. GREENE, A. M. TORRESSEN, S. VON MOLNAR, IBM, Thomas J. Watson Research Center, Yorktown Heights, NY 10598, J. G. BEDNORZ, K.A. MÜLLER, IBM Research, Ruschlikon, Switzerland -- We report specific heat measurements on the new high $T_c$ superconductors of the composition $La_{2-x}Ba_xCuO_{4-y}$, with $x \ll 1$ and $y > 0$. Polycrystalline samples with $x = .15$ show a resistivity drop of three orders of magnitude and a transition from Pauli paramagnetism to diamagnetism with an onset temperature between 30 – 35K.[1,2] The transition is complete by 10K and magnetic field studies suggest superconductivity of a percolative or granular nature. Our specific heat experiments indicate a large electron density of states but no evidence of a sharp jump near $T_c$ - consistent with the small Meissner signal observed (2% of complete flux expulsion) and the broad transition width. These measurements, along with x-ray and critical field results, will be analyzed for the possibility of high $T_c$ superconductivity in these new oxide materials.

1. J. G. Bednorz and K. A. Müller, Z. Phys. B **64**, 189 (1986).
2. J. G. Bednorz, M. Takashige and K. A. Müller, Europhysics Lett. Feb. 1987.

---

Fig. 1. Abstract of the only submitted talk about the high-Tc cuprates at the 1987 APS March meeting in New York. This contributed abstract was submitted in early December 1986 and was presented by the author (R. L. Greene) on the day before the famous "Woodstock of Physics" night session on the cuprates.

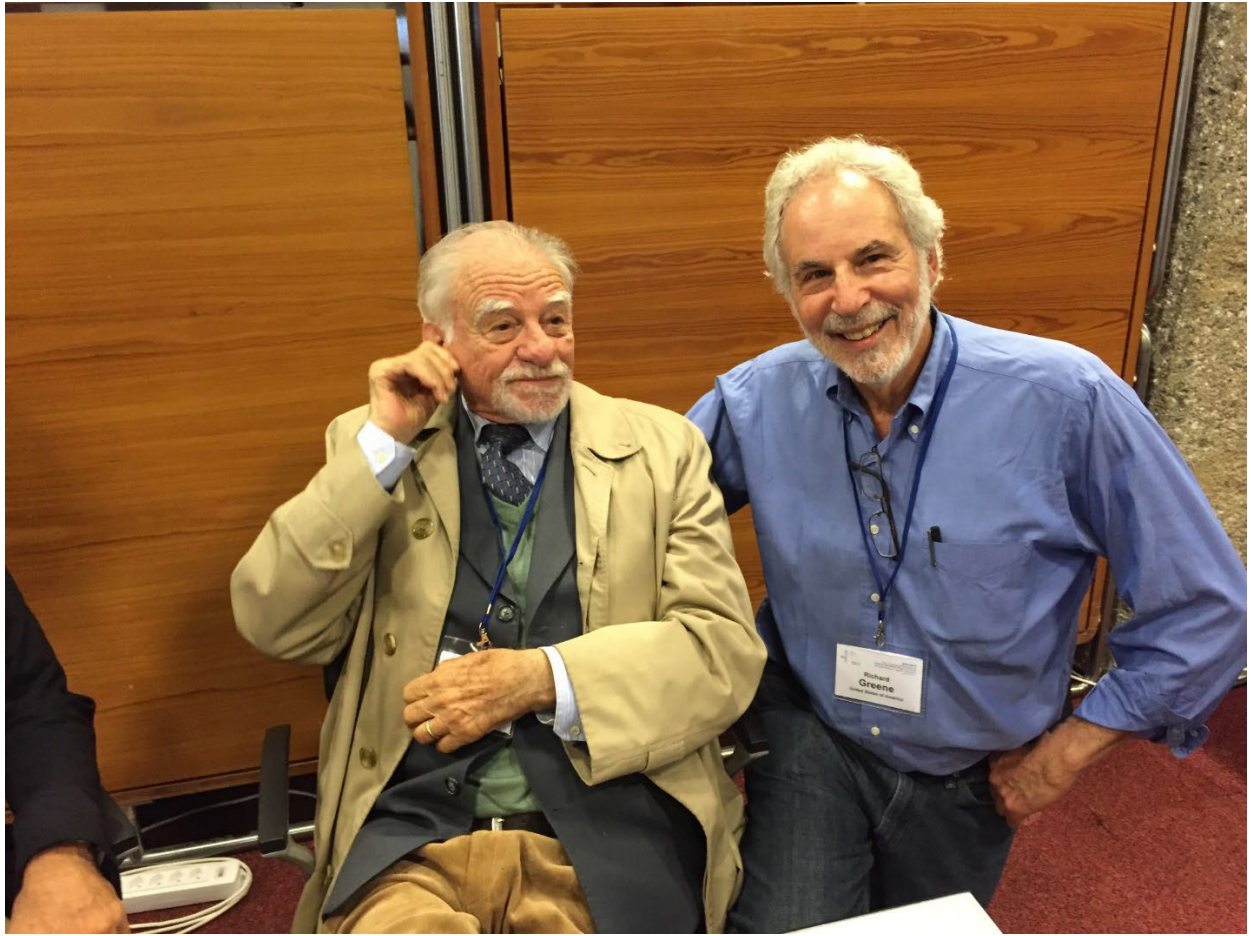

Fig. 2. K. Alex Müller and the author at the M2S conference in Geneva, Switzerland, July 2015.

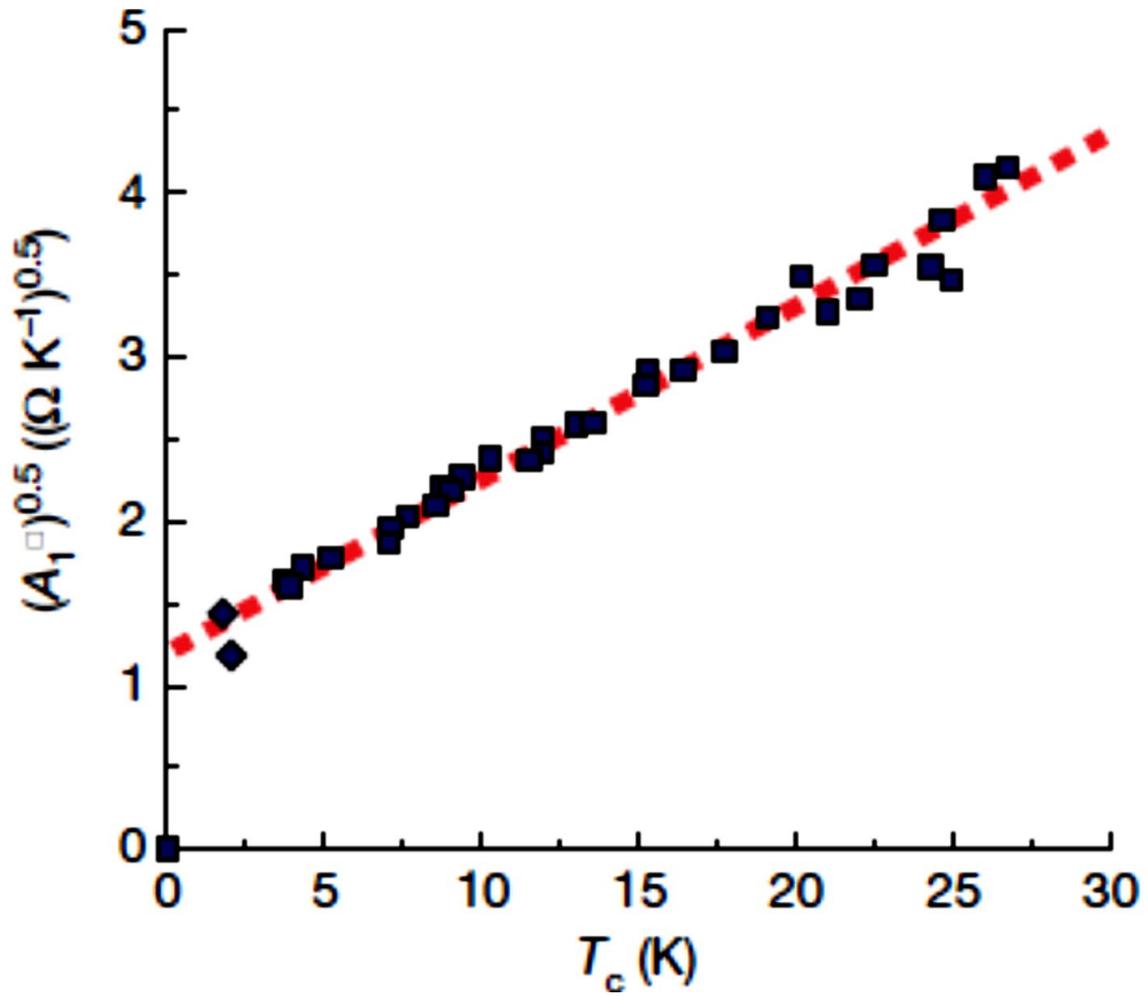

Fig. 3. Correlation between the slope of the normal state (H > Hc2) T-linear resistivity (A) and the superconductivity (Tc) in the electron-doped cuprate $La_{2-x}Ce_xCuO_4$ (LCCO). Data from ref. 23.

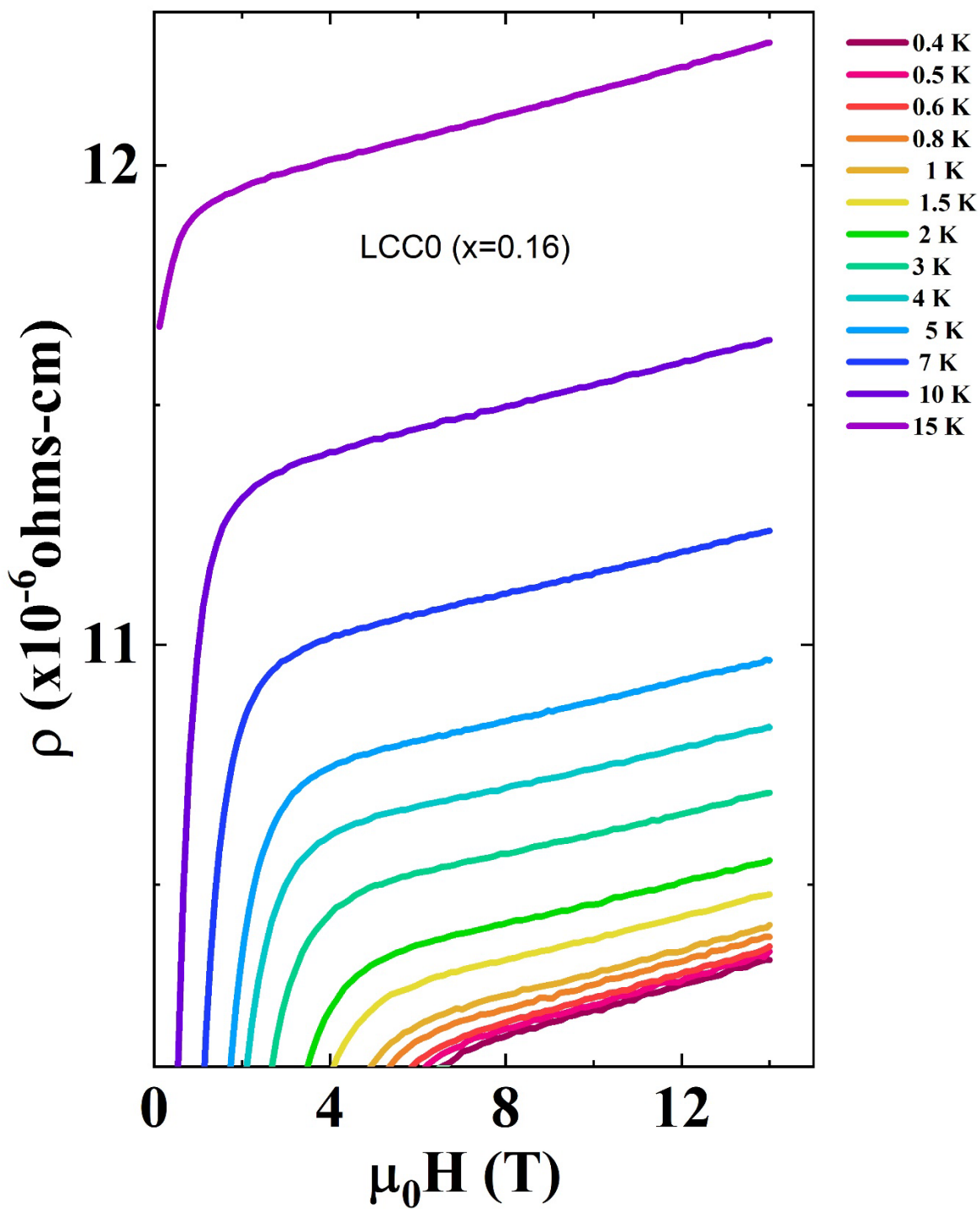

Fig. 4. Resistivity versus magnetic field (MR) at various temperatures for a thin film of the n-type cuprate LCCO with x = Ce = 0.16 doping. Adapted from ref. 28.

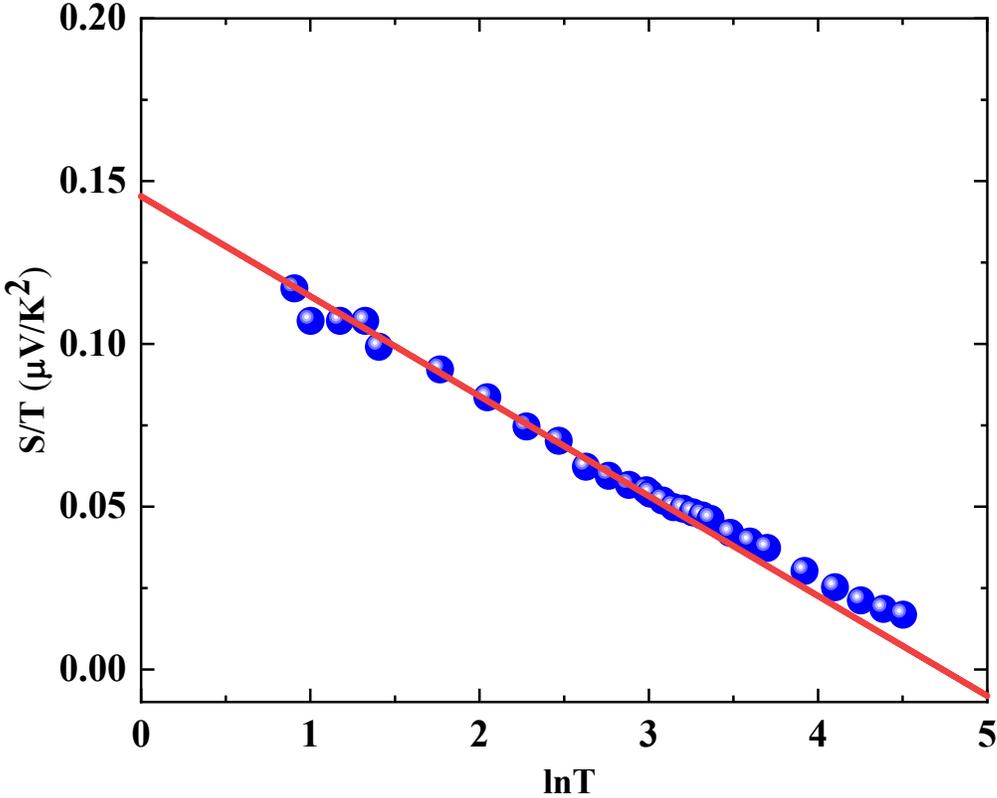

Fig. 5. Low temperature normal state (H = 11T) thermopower (S) plotted as S/T for LCCO with Ce = 0.16. Adapted from ref. 31.